\documentclass{article}

\usepackage[margin=1.2in]{geometry}
\usepackage{graphicx,amsmath,amsfonts,dsfont}
\usepackage{float}

\graphicspath{{fig/}}
\usepackage{natbib}
\usepackage{hyperref}
\usepackage{color,soul}

\begin{document}

\title{Kramnik vs Nakamura: A Chess Scandal}

\author{	
	\makebox[.4\linewidth]{Shiva Maharaj}\\
	\textit{ Chess Ed}\\
	\and
	\makebox[.4\linewidth]{Nick Polson\footnote{Nick Polson is Professor of Econometrics and Statistics at Chicago Booth: ngp@chicagobooth.edu }}\\
	\textit{  Booth School of Business}\\
	\textit{  University of Chicago}\\
	\and
	\makebox[.4\linewidth]{Vadim Sokolov\footnote{Vadim Sokolov is an Associate Professor at Operations Research at George Mason University: vsokolov@gmu.edu}}\\
	\textit{ Department of Systems Engineering }\\
	\textit{  and Operations Research}\\
	\textit{ George Mason University}\\
}

\maketitle

\begin{abstract}
\noindent We provide a statistical analysis of the recent controversy between Vladimir Kramnik (ex chess world champion) and Hikaru Nakamura. Hikaru Nakamura is a chess prodigy and a five-time United States chess champion. Kramnik called into question Nakamura's 45.5 out of 46 win streak in an online blitz contest at chess.com. We assess the weight of evidence using a priori  assessment of Viswanathan Anand and the streak evidence. Based on this evidence, we show that Nakamura has a  99.6 percent chance of not cheating. We study the statistical fallacies prevalent in both their analyses. On the one hand Kramnik bases his argument on the probability of such a streak is very small. This falls precisely into the Prosecutor's Fallacy. On the other hand, Nakamura tries to refute the argument using a cherry-picking argument. This violates the likelihood principle. We conclude with a discussion of the relevant statistical literature on the topic of fraud detection and the analysis of streaks in sports data.
\end{abstract}

\vspace{0.15in}

\section{Introduction}
We provide a statistical analysis of the recent controversy between Vladimir Kramnik (ex chess world champion) and Hikaru Nakamura. Hikaru Nakamura is a chess prodigy and a five-time United States chess champion. On November 2023 Kramnik, world chess champion from 2000 to 2007, insinuated in his chess.com blog post that Nakamura might have cheated while playing on the site. Kramnik called into question Nakamura's 45.5 out of 46 win streak in a 3+0 online blitz contest at chess.com. A 3+0 blitz game is a game where each player has 3 minutes to make all of their moves, with no after-move increment. Kramnik's argument is that the probability of such a streak is very small, thus we might have a case of cheating. Nakamura, on the other hand, claims that cherry-picking a sequence of 46 games out of more than 3500 he played on chess.com is not a fair approach to collect the data.The controversy has been widely discussed in the chess community and has been covered by major news outlets such as the BBC and the New York Times.

We assess the weight of evidence using a priori  assessment of Viswanathan Anand and the streak evidence. Based on this evidence, we show that Nakamura has a  99.6 percent chance of not cheating. We study the statistical fallacies prevalent in both their analyses. On the one hand Kramnik bases his argument on the probability of such a streak is very small. This falls precisely into the Prosecutor's Fallacy. On the other hand, Nakamura tries to refute the argument using a cherry-picking argument. This violates the likelihood principle. We conclude with a discussion of the relevant statistical literature on the topic of fraud detection and the analysis of streaks in sports data.


We start by addressing the argument of Kramnik which is based on the fact that the probability of such a streak is very small. This falls into precisely the Prosecutor's Fallacy, a statistical reasoning error that occurs when the probability of one event is confused with the probability of another related event. It's often associated with the misinterpretation of evidence and conditional probabilities. It assumes that the probability of innocence given the evidence is the same as the probability of the evidence given the innocence. This can lead to wrongly convicting innocent defendants.

Our approach mirrors the analysis of the Sally Clark case \citep{nobles2005misleading}, a British solicitor who was wrongfully convicted of murdering both of her children. The prosecution argued that the probability of two children dying of natural causes was 1 in 73 million. The fallacy was assuming that deaths were independent events. Other famous examples from the forensic science include cases of Mary Decker Slaney \citep{berry2004} and  O.J. Simpson \citep{good1996}. 

Denote by $G$ the event of being guilty, $I$ the event of innocence, and $E$ denote the evidence. In our case the evidence is the streak of wins by Nakamura. We are interested in probability of being innocent, given the winning streak $P(I\mid E)$. Kramnik's argument is that probability of observing the streak is very low, thus we might have a case of cheating. He simply says that $P(E\mid I)$ is low, thus Nakamura is not innocent. This is the prosecutor's fallacy
\[
P(I \mid E) \neq P(E \mid I).
\]
An intuitive example of the prosecutor's fallacy is the following
\[
P(\text{Play in NBA} \mid \text{Practice Hard}) \neq P(\text{Practice Hard} \mid \text{Play in NBA}).
\]
In this example, given we observed that a person practices hard playing basketball, we cannot make a conclusin that this person plays in NBA, the top basketball league in the world. The probability of playing in NBA given that a person practices hard $P(\text{Play in NBA} \mid \text{Practice Hard})$ is close to 0. At the same time, $P(\text{Practice Hard} \mid \text{Play in NBA})$ is close to 1. It is given, that all NBA players practice hard, but very few hard workers play in NBA.

The screenshot below shows the table that Kramnik presented at in interview on \textit{Levitov Chess World} YouTube channel (https://www.youtube.com/@LevitovChessWorld) he gave on November 27, 2023 \citep{levitovchessworld2023}.
\begin{table}[H]
\centering
\includegraphics[width=0.7\linewidth]{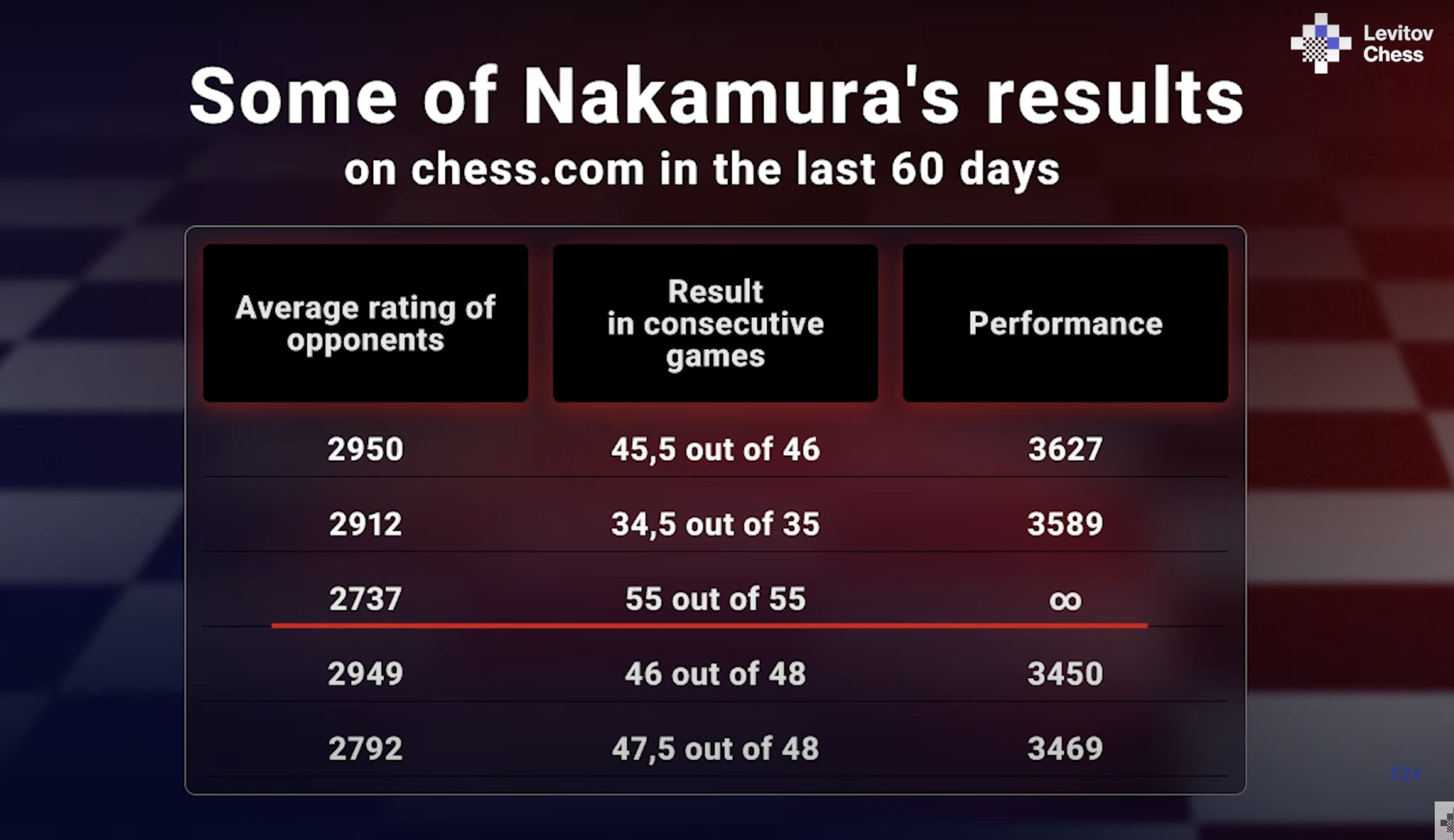}
\caption{Table Published by Kramnik. In this table Kramnik presents the implied performance of a player (Performance column) based on observed results and opponent's average ELO rating.}
\label{fig:kramnik-table}
\end{table}

The table shows the implied ELO rating of Nakamura (Performance column) given observed results and opponent's average ELO rating. The ELO rating, named after its creator Arpad Elo, is a method for calculating the relative skill levels of players in zero-sum games such as chess. The table shows that the implied ELO rating of Nakamura is 3627 given 45 win streak, which is significantly higher than his current rating of 3300.

Although, it is not clear how exactly the ``Performance" column was constructed, the analysis of Kramnik follows the approach that uses $P(E\mid I)$ rather than $P(I\mid E)$ (prosecutor's fallacy). As we show in the next section, based on our analysis, the likelihood of a streak of 45 wins is $P(E\mid I) = 0.0286$, while the posterior probability of not cheating is $P(I\mid E) = 0.9965$. The prosecutor's fallacy can lead to an overestimation of the strength of the evidence and may result in an unjust conviction. 

Further, Kramnik's calculation neglects other relevant factors, such as the prior probability $P(G)$ of cheating.  At the top level of chess, prior probability of cheating $P(G)$ is small due to reputational effects. According to a recent statement by Viswanathan Anand, the probability of cheating is around $1/10000$. Viswanathan Anand is an Indian chess grandmaster and a five-time world champion (2007–2013). He is widely regarded as one of the greatest chess players of all time. Anand's prior probability of cheating is based on personal assessment of the proportion of cheaters in the chess community.
\begin{figure}[H]
	\centering
	\includegraphics[width=0.7\textwidth]{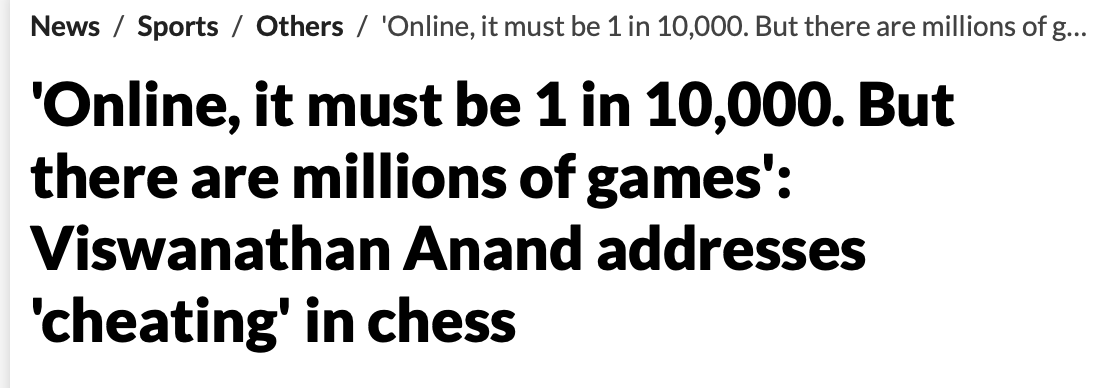}
	\caption{Anand's prior. Source: \href{https://www.hindustantimes.com/sports/others/online-it-must-be-1-in-10-000-but-there-are-millions-of-games-viswanathan-anand-addresses-cheating-in-chess-101668249442048.html}{Hindustan Times.}}
\end{figure}

The next section provides calculations that use the prior information of cheating and applies Bayes rule to calculate the probability of cheating given the observed 45 win streak. We assess that Nakamura has 99.6 percent chance of not cheating given Anand's prior assumption. Bayes rule uses two inputs to calculate the probability of cheating given the evidence. The first input is the prior odds of guilt. The second is the Bayes factor, which depends on evidence. Given the subjective nature of the prior odds, we provide a sensitivity analysis to this input inherent in the analysis of the Nakamura-Kramnik controversy. Nakamura is one of the world's best 3+0 online blitz players, second only to Magnus Carlsen, the current world champion. We show that this winning streak, given he played against lower rated players, does not provide strong evidence of cheating. We also discuss relevant research on the topic of streaks in sports.



\section{Bayesian Analysis}\label{sec:island}
The first component of our analysis is the prior information, which is given via prior odds of being guilty $O(G) = P(I)/P(G)$. Here $P(I)$ and $P(G) = P(\text{not }I) = 1-P(G)$ are the prior probabilities of innocence and guilt respectively. The second term is the Bayes factor, which is the ratio of the probability of the evidence under the innocence  hypothesis to the probability of the evidence under the guilt hypothesis. The Bayes factor is given by
\[
	L(E\mid G) = \frac{P(E\mid I)}{P(E\mid G)}.
\]

In the odds from of the Bayes rule, product of the Bayes factor and the prior odds is the posterior odds of guilt, given the evidence. The posterior odds of guilt is given by
\[
	O(G\mid E) = O(G) \times L(E\mid G).
\]
Anand stated that there one cheater out of 10000 players. In general, when our prior assumption is that out of $N+1$ players, there is one cheater, we have $P(G) = 1/(N+1)$ and $P(I) = N/(N+1)$, the  prior odds of guilt are
\[
	O ( G )  = \dfrac{N/(N+1)}{1/(N+1)} = N.
\]
Finally, if we denote the likelihood of innocence, $P(E\mid G)$ as $p$, the Bayes factor is given by
$$
\frac{P(E\mid I)}{P(E\mid G)} = \dfrac{p}{1} = p.
$$ 
Thus, the Bayes rule states that the posterior odds of guilt is a function of $p$ and $N$ and is given by
\[
O(G\mid E) = Np.
\]
There are two numbers we need to estimate to calculate the odds of cheating given the evidence, namely the prior probability of cheating given via $N$ and the probability of a streak $p = P(E\mid I)$. 

If a tie is treated as a no-win event, the outcomes of each game are binary with the outcomes win and no-win. There are multiple ways to calculate the probability of a streak. We can use the binomial distribution, the negative binomial distribution, or the Poisson distribution. Given, we know the number of games Kramnik used for his analysis, the binomial distribution is a more natural choice compared to Poisson. The probability of $k$ wins out of $m$ games is given by 
\[
p = 	P(E\mid I) = \binom{m}{k} q^k (1-q)^{m-k}.
\]
Here $q$ is the probability of winning a single game. Thus, for a streak of 45 wins, we have $k = 45$ and $m = 46$.

The individual game win probability $q$ is calculated from the ELO rating difference between the players.  The ELO rating of Hikaru is 3300 and the average ELO rating of his opponents is 2933, according to Kramnik. The difference of 366 corresponds to the odds of winning of $w = 10^{366/400} = 10^{0.92} = 8.23$. Using this estimate, the average probability of winning a single game is then $q = w/(1+w) = 0.8916$. Nakamura reported similar estimate,  on his YouTube channel. Plugging in the values for $k$, $m$, and $q$ we calculate the probability of a 45 wins in 46 games to be  $p = 0.0286$.



Under, Anand's prior odds of cheating, we have  $N = 10000$.  By Bayes rule, the posterior odds of cheating is $O(G \mid E) = Np = 286$.  Therefore, the probability of cheating is 
$$P(G\mid E) = 1/(1+O(G\mid E)) = 0.003491,$$ 
and the probability of innocence is $P(I \mid E) = 1 - 0.0035 = .9965$.

Note, that the probability of being innocent depends on the prior probability of cheating. For completeness, we perform sensitivity analysis and also get the odds of not cheating for a different value of $N$. One can argue that Anand is very conservative in his assessment of the prior probability of cheating. Thus, we  calculate the probability of innocent for assuming that chewing is more prevailing than Anand is thinking and assume $N = 500$. Under this assumption, we get
\[
	P(G\mid I) = \frac{1}{1 + 500 \times 0.0285} = 0.065, P(I\mid E) = 1 - P(G\mid I) = 0.935.
\]

\begin{figure}[H]\label{fig:n-p}
	\centering
	\includegraphics[width=1\textwidth]{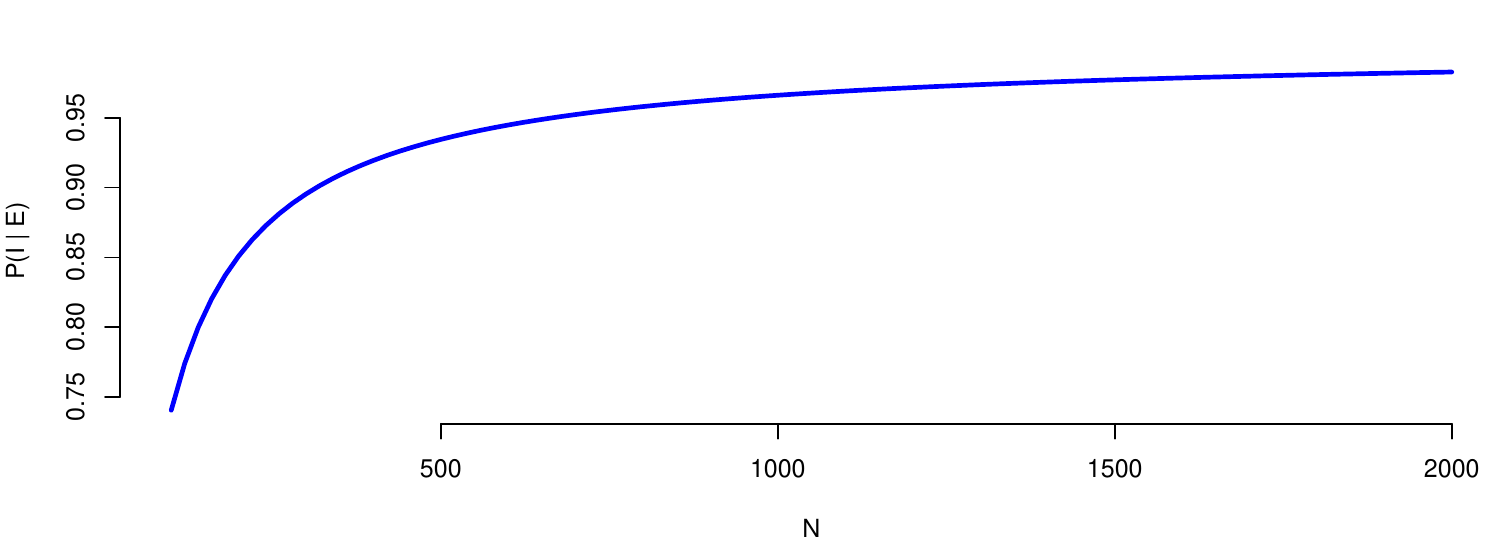}
	\caption{Relationship between the prior $N$ and the posterior probability of not cheating with likelihood of strike $p = P(E\mid I) = 0.0286.$ }
\end{figure}

Next, we plot the posterior probability of not cheating given the evidence of a streak of 45 wins in a row for different values of $N$, ranging between 100 and 2000. The plot is shown in Figure \ref{fig:n-p}. The plot shows that the posterior probability of cheating is very sensitive to $N$. The posterior probability of not cheating is "flattening out" when $N$ is greater than 1500.

There are several assumptions we made in this analysis. 

\begin{enumerate}
	\item Instead of calculating game-by-game probability of winning, we used the average probability of winning of $q$. If we use the game-by-game probability of winning, the probability of a streak would be given by Poisson binomial distribution which has mass function proportional to the product of the probabilities of winning each game. Under our assumption the mass function is proportional to the average probability of winning raised to the power of the number of games. Our assumption decreases the likelihood and slightly shifts posterior odds in favor of not cheating. Due to the following fact (known as Jensen's inequality). Given a sequence of random draws $q_1,\ldots,q_m$
	\[
	\prod_{i=1}^{m}q_i < q^m, ~ \text{where} ~  q = \mathrm{Average}(q_1,\ldots,q_m)= \frac{1}{m}\sum_{i=1}^{m}q_i.
	\]
	However, the closer $q$ is to 1, the smaller the difference between the left and right-hand sides of the inequality. Thus, use of average probability is a reasonable assumption given the fact that Nakamura is a much stronger player than his opponents and the difference in the ELO ratings is large with $q=0.8916$ being close to 1. Empirically, if we generate $q_i \sim U(0.85,0.9),~i=1,\ldots,46$, then $\prod_{i=1}^{m}q_i / q^m$ will be around 0.995. 

	\item Further, there is some correlation between the games, which also shifts the posterior odds in favor of not cheating. For example, some players are on tilt. Tilt happens when player us frustrated as a result of loosing several games and adopts a suboptimal strategy, usually resulting in the player becoming overly aggressive. Simply speaking, given they lost first game, they are more likely to lose the second game.
	\item There are many ways to win 3+0 unlike in classical chess. For example, one can win on time. If you win on time it means that your opponent did not move before the time limit. We argue that probability of winning calculated from the ELO rating difference is underestimated. 
\end{enumerate}




\section{Discussion}


In sum, our Bayesian analysis of the Nakamura-Kramnik controversy assesses that Nakamura has  99.9 percent chance of not cheating given Anand's prior assessment of the odds of cheating.
The case is interesting as it highlights many of the statistical fallacies that can arise in data analysis. Kramnik bases his argument on the fact that the probability of such a streak is very small. This falls into precisely the Prosecutor's Fallacy. See \cite{balding1994prosecutor}. Nakamura tries to refute the argument using a cherry-picking argument and mentions that Kramnik cherry-picked a sequence of 46 games out of more than 3500 he played on chess.com. This falls into a violation of the likelihood principle. See \cite{berger1988likelihood}. The likelihood principle \citep{edwards1963} is a fundamental concept in Bayesian statistics that states that the evidence from an experiment is contained in the likelihood function. It implies that the rules governing when data collection stops are irrelevant to data interpretation. It is entirely appropriate to collect data until a point has been proven or disproven, or until the data collector runs out of time, money, or patience. As Edwards-Lindman-Savage say it: ``Often evidence which, for a Bayesian statistician, strikingly supports the null hypothesis leads to rejection of that hypothesis by standard classical procedures.'' 

Another fallacy is Cromwell's rule. Cromwell's rule,  states that predictions should be wary of assigning a prior probability of 0 (impossible) or 1 (certain) to anything except to statements that must be logically true. Essentially, $P(G)=0$ implies $P(G\mid E) =0$ for any evidence $E$.  GM Benjamin Finegold in his tweet on November 20, 2023, states that the probability of Nakamura's streak is zero.
\begin{figure}[H]
\centering
\includegraphics[width=0.7\linewidth]{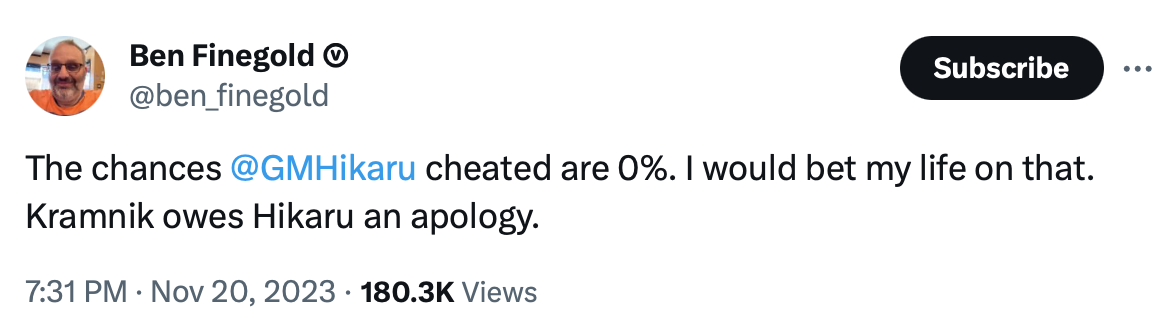}
\caption{Screenshot of Ben Finegold's Tweet from November 20, 2023.}
\label{fig:}
\end{figure} 

In general, analysis of streaks is statistical area of interest. From a theoretical perspective, \cite{aldous1989} proposes the use of a Poisson clumping heuristic to analyze the length of streaks. From an empirical perspective, there have been many interesting applications, for example, in baseball (see \cite{albert1993a}). Analysis of streaks in chess is a fruitful area for future study.  

One should always keep in mind the idea that in the light of new evidence one should allow the possibility of updating our believes. As Kaynes said, "Sir, if the facts change, I change my opinion".

\vspace{2em}
\noindent
\textbf{Acknowledgement}: We would like to thank the associate editor for the helpful comments and suggestions.

\bibliographystyle{plainnat}
\bibliography{ref,NakamuraKramnik}

\end{document}